# Adoption and implication of the Biased-Annotator Competence Estimation (BACE) model into COVID-19 vaccine Twitter data: Human annotation for latent message features

Luhang Sun[1], Yun-Shiuan Chuang[23], Yibing Sun[1], and Sijia Yang[1]

## Abstract

Traditional quantitative content analysis approach (human coding method) has weaknesses, such as assuming all human coders are equally accurate once the intercoder reliability for training reaches a threshold score. We applied the Biased-Annotator Competence Estimation (BACE) model (Tyler, 2021), which draws on Bayesian modeling to improve human coding. An important contribution of this model is it takes each coder's potential biases and reliability into consideration and treats the "true" label of each message as a latent parameter, with quantifiable estimation uncertainties. In contrast, in conventional human coding, each message will receive a fixed label without estimates for measurement uncertainties. In this extended abstract, we first summarize the weaknesses of conventional human coding; and then apply the BACE model to COVID-19 vaccine Twitter data and compare BACE with other statistical models; finally, we discuss how the BACE model can be applied to improve human coding of latent message features.

[1] School of Journalism and Mass Communication, University of Wisconsin–Madison, Madison, WI, USA.
[2] Department of Psychology, University of Wisconsin–Madison, Madison, WI, USA.
[3] Department of Computer Sciences, University of Wisconsin–Madison, Madison, WI, USA.

**Corresponding Author**: Sijia Yang, University of Wisconsin–Madison, Vilas Communication Hall, 821 University Avenue, Madison, WI 53706, USA.
Email: syang84@wisc.edu

**Conventional human annotation: traditional approach and its weaknesses**

In the communication discipline, it is common for researchers to measure latent message attributes by training a group of people (usually two or three undergraduate students) as human coders. Researchers in social science (e.g., Gheyle & Jacobs, 2017; Hoover et al., 2020) usually call this the human annotation (human coding) approach toward quantitative content analysis (QCA, Gheyle & Jacobs, 2017; Rourke & Anderson, 2004). Traditional coding procedure starts with a codebook based on research questions. Coders will be trained based on the concepts from the coding scheme. By randomly subsetting a small sample of the data, coders will code the same content and intercoder reliability scores are calculated by indices, such as Cronbach's alpha, Cohen's kappa, and Krippendorff's alpha (Gheyle & Jacobs, 2017; Hayes & Krippendorff, 2007). If the score is lower than the expected threshold (E.g., 0.7), coders need to gather together to discuss and refine the coding scheme, and the codebook is sometimes revised accordingly as well. Once the intercoder reliability score hits the threshold, coders are assumed to have a similar understanding of the concepts in the coding scheme and are ready to code the rest of the raw data separately based on researchers' goals.

However, in reality, it is common that repetitive training processes may not produce an ideal improvement in the intercoder reliability scores. For example, past research has shown that extensive training of human coders fails to improve the reliable coding of moral appeals in news stories, and better results are obtained through crowdsourcing (Weber et al., 2018). Moreover, even if the researcher's selected reliability metric passes the conventional threshold, human coders are still susceptible to measurement errors and such errors should be incorporated into the final "labels" rather than assumed away. In summary, the traditional human coding approach has three main weaknesses (Tyler, 2021): first, the traditional approach assumes that all coders will make no mistakes once their intercoder reliability reaches the threshold. Second, the traditional approach assumes that all coders will be equally correct and unbiased. Third, the traditional approach assumes that the latent message attributes are labeled without measurement uncertainties.

With all the weaknesses of the traditional approach in mind, in the current study, we adopt a new Bayesian statistical model called Biased-Annotator Competence Estimation (BACE) from Tyler (2021). As a case study, we share our experiences applying this method to increase the trustworthiness of coded valences of tweets related to COVID-19 vaccines. This study first uses the traditional human annotation approach to create a latent variable, valence toward COVID-19 vaccines; then adopts the BACE model to improve the reliability of our coded data. We also compared this new Bayesian approach with more conventional statistical models. Although BACE does not overperform other models in our case study, we noted several strengths of BACE and encourage its adoption in future research on latent message features. These strengths include the following: BACE admits that coders may have different levels of biases toward certain labels and provides an estimate of each coder's competence parameters and bias probabilities, which are useful for identifying coders' competence and bias levels, respectively; it also produces an uncertainty measure for the estimated "true" label for each coded message.

**Adoption of Biased-Annotator Competence Estimation (BACE) in valence prediction of COVID-19 vaccine Twitter data**

To solve the human coding accuracy issues, Hovy et al. (2013) introduced their unsupervised predicting model Multiple Annotator Competence Estimation (MACE), to evaluate the trustworthiness of coders and predict the true latent labels. Developed from MACE, the BACE model (Tyler, 2021) not only assigns each coder a competence parameter based on their trustworthiness but also considers and allows biased coders (e.g., some coders may be inclined to code certain labels unconsciously) to exist in the coding team. The BACE model treats each data point's true label as a random variable with a distribution, while each coder's answer is treated as a draw from this distribution with noises. The true labels are predicted based on the coders' competencies and are presented as predicted probabilities as outputs. Each coder's competence parameter is a parameter of their accuracy based on pairwise agreement and disagreement between coders: when the competence parameter is higher, the coder is more likely to code the true label.

***Data and coding procedure***

The data we use for this study is a COVID-19 vaccine Twitter dataset. We collected all tweets related to COVID-19 vaccines from Twitter Academic API 2.0 using a list of keywords relevant to COVID-19 vaccines (see Appendix A). The time frame is from December 1, 2020 to June 30, 2022. For the exact human coding task in the current study, we extracted a random subset of tweets ($N_{Total\ coded}$ = 9,373) for human coding. Since it is recommended to have a dataset coded by at least three coders by the BACE model, we extracted our three-coder subset ($N_{3\text{-}coder\ coded}$ = 3,021).

For the traditional human coding approach, our coding scheme includes two variables (relevancy and valence) with two steps: first validating if the tweet is relevant to COVID-19 vaccines or not, as the keyword-based data pulling may include false positives or noises, and then identifying the valence, if the tweet is generally positive, negative, or unclear toward COVID-19 vaccines. We followed the traditional coding procedure starting from coder training and calculated intercoder reliabilities using ReCal (Freelon, 2010, 2013); however, the intercoder reliability was lower than the conventional threshold. For our first variable, COVID-19 vaccine relevancy, the intercoder reliability reached 0.88 for Krippendorff's alpha. However, we recognized that re-training did not help improve the intercoder reliability for the valence variable—Krippendorff's alpha was always lower than 0.7.

Table 1 shows the intercoder reliability scores for COVID-19 vaccine valence for this 3-coder-coded subset ($N_{3\text{-coder valence}}$ = 1,659). The total volume of the valence subset is lower than the three-coder subset ($N_{3\text{-coder coded}}$ = 3,021) because we removed all the irrelevant tweets in Step 1. And all data points in this 3-coder-coded subset ($N_{3\text{-coder valence}}$ = 1,659) for valence are relevant to COVID-19 vaccines. Krippendorff's alpha for the valence variable was only 0.55 (Table 1). It might be because there were many tweets talking about COVID-19 vaccines with ambiguity and sarcasm, preventing coders from reaching a high agreement. Therefore, we adopted the BACE model to re-estimate valence while explicitly quantifying coders' potential biases and competence.

***Results***

We used the R package of BACE to predict the true labels of COVID-19 vaccine valence in our subset. To compare BACE with other statistical models, we also tried a majority model and a Dawid-Skene (DS) model (Dawid & Skene, 1979) in the current study. The majority model predicts the label based on the majority vote of all coders. If there are ties in the votes (e.g., each of the three coders selected one of the three labels), the model breaks the tie based on the categorical distribution of the coders' labels. The DS model was designed and widely used for diagnostic medicine. Different from BACE, the DS model is flexible and can equip each coder-label with a competence parameter in the matrix of all potential combinations between coders (e.g., Coder 1, Coder 2, and Coder 3) and labels (e.g., Label 1, Label 2, and Label 3).

Since the true label is a random unknown variable that is predicted by BACE, our study team also coded a small subset of randomly sampled tweets as the "ground truth" against which to evaluate the prediction accuracy for each model (see Table 2). We divided the true labels coded by experienced researchers into two subsets, "ambiguous set" and "clear set", based on coders' agreement and disagreement. As mentioned above, the biggest difficulty that impeded coders from reaching a high agreement for the valence variable is because of the ambiguity and sarcasm of many tweets discussing COVID-19 vaccines. Therefore, experienced researchers identified and divided the assessment criteria into the two subsets in Table 2. Comparing with each other of the three statistical models and Krippendorff's alpha (0.55) in Table 1, we can find first, all three models perform better than the traditional approach, which is directly treating coders' answers as the true labels. In this case study, BACE reached comparable decisions as the majority model and the DS model but did not outperform the other two.

**Discussion and looking forward**

Although BACE does not overperform the two other statistical models in the current case study, we believe there still are many benefits to introducing and adopting the BACE model into the conventional human coding approach. The biggest advantage of BACE is that it takes coders' unconscious biases into consideration and can provide both competence parameters *($\beta$)* and bias probabilities *($\gamma$)* from all coders (see Table 3) for researchers to better understand coders' performance separately rather than treating all coders as equally accurate and trustworthy without considering their unconscious biases. Furthermore, since BACE treats the latent message "label" as a random variable with a distribution, measurement uncertainties can be estimated straightforwardly in the Bayesian modeling framework. Analogously, BACE can produce a 95% "confidence interval" for each message's label rather than assuming it is measured without error. Therefore, it is beneficial for researchers to adopt BACE into the human annotation process, especially for latent message features. BACE can also be applicable to crowdsourcing annotation, which involves more annotators instead of two or three coders in the traditional approach. The nature of "a large number of minimally trained annotators" (Hopp et al., 2021) of the crowdsourcing approach can benefit from BACE because it provides both competence parameters and bias probabilities for all annotators.

To conclude, in the present study, we adopt a new Bayesian approach, BACE, to the human annotation process, and encourage researchers to apply BACE to future studies for latent message features because BACE not only provides an estimate of each coder's competence and bias, but also produces an uncertainty measure for the estimated "true" label for each coded message.

## References

Dawid, A. P., & Skene, A. M. (1979). Maximum likelihood estimation of observer error-rates using the EM algorithm. *Journal of the Royal Statistical Society: Series C (Applied Statistics), 28*(1), 20-28.

Freelon, D. (2010). ReCal: Intercoder reliability calculation as a web service. *International Journal of Internet Science, 5*(1), 20-33.

Freelon, D. (2013). ReCal OIR: Ordinal, interval, and ratio intercoder reliability as a web service. *International Journal of Internet Science, 8*(1), 10-16.

Gheyle, N., & Jacobs, T. (2017). Content Analysis: a short overview. *Internal research note*, 1-17.

Hayes, A. F., & Krippendorff, K. (2007). Answering the call for a standard reliability measure for coding data. *Communication methods and measures, 1*(1), 77-89.

Hoover, J., Portillo-Wightman, G., Yeh, L., Havaldar, S., Davani, A. M., Lin, Y., ... & Dehghani, M. (2020). Moral foundations twitter corpus: A collection of 35k tweets annotated for moral sentiment. *Social Psychological and Personality Science, 11*(8), 1057-1071.

Hopp, F. R., Fisher, J. T., Cornell, D., Huskey, R., & Weber, R. (2021). The extended Moral Foundations Dictionary (eMFD): Development and applications of a crowd-sourced approach to extracting moral intuitions from text. *Behavior research methods, 53*(1), 232-246.

Hovy, D., Berg-Kirkpatrick, T., Vaswani, A., & Hovy, E. (2013, June). Learning whom to trust with MACE. In *Proceedings of the 2013 Conference of the North American Chapter of the Association for Computational Linguistics: Human Language Technologies* (pp. 1120-1130).

Rourke, L., & Anderson, T. (2004). Validity in quantitative content analysis. *Educational technology research and development, 52*(1), 5-18.

Tyler, M. (2021). *Essays in Political Methodology*. Stanford University.

Weber, R., Mangus, J. M., Huskey, R., Hopp, F. R., Amir, O., Swanson, R., ... & Tamborini, R. (2018). Extracting latent moral information from text narratives: Relevance, challenges, and solutions. *Communication Methods and Measures*, *12*(2-3), 119-139.

**Table 1.** Intercoder reliability scores for coding valence for the three-coder coded subset

| Intercoder reliability | Value |
|---|---|
| N coders | 3 |
| N cases | 1659 |
| N decisions | 4977 |
| *Average Pairwise Percent Agreement* | 71.57% |
| Pairwise Agreement cols 1 & 3 | 0.706 |
| Pairwise Agreement cols 1 & 2 | 0.711 |
| Pairwise Agreement cols 2 & 3 | 0.730 |
| *Fleiss' Kappa* | 0.550 |
| FK observed agreement | 0.716 |
| FK expected agreement | 0.368 |
| *Average Pairwise Cohen's Kappa* | 0.550 |
| Pairwise CK cols 1 & 3 | 0.538 |
| Pairwise CK cols 1 & 2 | 0.542 |
| Pairwise CK cols 2 & 3 | 0.571 |
| *Krippendorff's alpha* | 0.550 |

**Table 2.** Model accuracy table

| Model | Ambiguous set | Clear set |
|---|---|---|
| BACE model | 0.59 | 0.98 |
| Majority model | 0.68 | 0.98 |
| DS model | 0.68 | 0.98 |

**Table 3.** The parameters of competence and biases for all three coders

| | Competence parameter (*beta*) | Bias probability (*gamma*) | | |
|---|---|---|---|---|
| | | Negative | Neutral or Unclear | Positive |
| Coder 1 | 0.784 | 0.085 | 0.693 | 0.222 |
| Coder 2 | 0.728 | 0.192 | 0.614 | 0.194 |
| Coder 3 | 0.755 | 0.177 | 0.649 | 0.174 |

**Appendix A. The keyword list for COVID-19 vaccine Twitter data collection**

vaccine OR vaccines OR vaccination OR vaccinations OR vaccinate OR vaccinated OR vax OR vaxx OR vaxxx OR vaxxed OR covax OR shot OR shots OR dose OR doses OR covidvaccine OR covid19vaccine OR coronavaccine OR coronavirusvaccine OR covaxin OR mrna OR nvic OR booster OR boosters OR pfizer OR moderna OR gamaleya OR "oxford-astrazeneca" OR astrazeneca OR cansino OR "johnson & johnson" OR "j&j" OR "j & j" OR "vector institute" OR novavax OR sinopharm OR sinovac OR "bharat biotech" OR janssen OR cepi OR biontech OR sputnikv OR bektop OR zfsw OR nvic OR pfizerbiontech OR "biontechvaccine" OR "warp speed" OR "delta variant" OR oxfordvaccine OR pfizervaccine OR pfizercovidvaccine OR modernavaccine OR modernacovidvaccine OR biotechvaccine OR biotechcovidvaccine OR biontechvaccine OR biontechcovidvaccine OR bektopvaccine OR simopharmvaccine OR johnsonvaccine OR janssenvaccine OR azvaccine OR astrazenecacovidvaccine OR astrazenecavaccine OR thisisourshot OR vaxhole OR notocoronavirusvaccines OR getvaccinated

**Appendix B. Human Coding Scheme**

Undergraduate coders were trained for coding two variables: relevancy of COVID-19 vaccines and valence toward COVID-19 vaccines. Coders were required to code and make judgments in Qualtrics.

***Step 1. Relevancy: Is this tweet relevant to COVID-19 vaccines or not? (1=YES, 0=NO)***

<table>
<tr><td>Code 1</td><td>Tweets relevant to the COVID-19 vaccines.</td></tr>
<tr><td>Code 0</td><td>The majority of the data should be relevant to COVID-19 vaccines because some tweets in the dataset may be irrelevant to the COVID-19 vaccines.<br>Examples:<br>1. Only mentioning other vaccines, like MMR, HPV, flu shot, Smallpox vaccine, etc., but not mentioning the COVID vaccine;<br>2. Other news or comments related to the pharmaceutical companies' other products, like Johnson & Johnson or Pfizer<br>a. <i>E.g., "I feel sorry for women who thought the misery of having their husband humping them has to continue because @pfizer's #Viagra. Just when she thought the misery was over, he gets a prescription."</i><br>3. Tweets with irrelevant hashtags to get traffic on Twitter.</td></tr>
</table>

If Step 1 is 1, then move to Step 2 to code its valence.

If Step 1 is 0, stop coding and move to the next one.

***Step 2. Valence: What is the valence of this tweet, positive/pro-vaccine, negative/anti-vaccine, or neural? (1=Positive and pro-vaccine, 2=Negative and anti-vaccine, 0=Neutral or unclear)***

<table>
<tr><td rowspan="2">Code 1</td><td>Generally positive attitudes toward the COVID-19 vaccines</td></tr>
<tr><td>Examples:<br>1. Refutation (Counter-hesitancy) of the statements/people/behaviors of anti-COVID vaccines.<br>2. Direct promotion of the COVID vaccine.</td></tr>
</table>

<table>
<tr><td rowspan="2">Code 2</td><td>Generally negative attitudes toward the COVID-19 vaccines</td></tr>
<tr><td>Examples:<br>1. Expressing hesitancy and undecided sentiment toward the COVID vaccine, or preferring to wait for a longer time to get vaccinated.<br>2. Anti-vaccine: expressing strong opposition toward the COVID vaccine.</td></tr>
<tr><td rowspan="2">Code 0</td><td>Neutral or unclear valence</td></tr>
<tr><td>Examples:<br>1. Other tweets that do not express explicit valence should be coded 0.<br>2. Sarcasm should be coded 0.</td></tr>
</table>